\documentclass[12pt]{article}
\usepackage{amsfonts,amsmath}
\usepackage{amssymb}

\newcommand{\dr}{{{\rm d}}}

\makeatletter \@addtoreset{equation}{section} \makeatother

{\vspace{3mm} }

\def\al{\alpha}

\def\*{\star}
\def\e{\mathbf{e}}
\def\E2{\mathbf{E}}

\newcommand{\be}{\begin{equation}}
\newcommand{\ee}{\end{equation}}
\newcommand{\bee}{\begin{eqnarray}}
\newcommand{\beee}{\begin{array}}
\newcommand{\eee}{\end{eqnarray}}
\newcommand{\eeee}{\end{array}}
\newcommand{\gb}{\beta}
\newcommand{\gga}{\gamma}

\newcommand{\gd}{\delta}
\newcommand{\gl}{\lambda}

\newcommand{\go}{\omega}

\newcommand{\dal}{\dot \alpha}
\newcommand{\dgb}{\dot \beta}
\newcommand{\dgga}{\dot \gamma}

\begin{document}
\begin{flushright}
FIAN/TD/10-2026\\
\end{flushright}

\vspace{0.5cm}
\begin{center}
{\large\bf Self-dual gravity from higher-spin theory}

\vspace{1 cm}

\textbf{V.E.~Didenko and A.V.~Korybut}\\

\vspace{1 cm}

\textbf{}\textbf{}\\
 \vspace{0.5cm}
 \textit{I.E. Tamm Department of Theoretical Physics,
Lebedev Physical Institute,}\\
 \textit{ Leninsky prospect 53, 119991, Moscow, Russia }
\par\end{center}

\begin{center}
\vspace{0.6cm}
e-mails: didenko@lpi.ru, akoribut@gmail.com \\
\par\end{center}

\vspace{0.4cm}

\begin{abstract}
\noindent Higher-spin symmetry is known to mix lower-spin fields with higher-spin fields, creating a complex interaction picture where no closed finite field sector is expected to exist for dimensions greater than three. By studying the self-dual part of higher-spin interaction vertices in four dimensions, we show that gauge fields of spins greater than two can be consistently set to zero. In this case, the fields with helicities $-2\leq\lambda\leq 0$ form a closed sub-sector and also act as sources for positive helicities. For these lower spin fields, we identify their equations of motion. In particular, we show that self-dual gravity with a cosmological constant emerges as a unique rigid part of higher-spin interactions. Notably, its equations have a form that incorporates the Moyal star product, which is essential for generating the higher-spin algebra. Therefore, we demonstrate that self-dual gravity can be derived from higher-spin symmetries.

\end{abstract}

\section{Introduction}

Higher Spin (HS) gauge theory; see reviews \cite{Vasiliev:1999ba, Sorokin:2004ie, Bekaert:2004qos, Didenko:2014dwa, Ponomarev:2022vjb}, is a framework for interacting gauge fields with spins $s\geq 1$  (which includes gravity) and matter fields with spins  $s = 0$  and  $s = 1 / 2$. The interactions are governed by higher-spin symmetry \cite{Fradkin:1986ka, Eastwood:2002su, Vasiliev:2003ev}, which extends the conventional space-time Poincar\'{e} algebra by introducing an infinite number of generators $T_s$ parameterized by spin $s$. The commutation relations can be schematically expressed as:
\begin{equation}\label{HS:com}
    [T_{s_1}, T_{s_2}]=T_{s_1+s_2-2}+T_{s_1+s_2-4}+\dots\,.
\end{equation}
This equation indicates that for spins greater than two, the algebra acquires infinitely many generators. The standard no-go arguments due to Coleman and Mandula \cite{Coleman:1967ad} and Weinberg \cite{Weinberg:1964ew} that highly constrain nontrivial interactions in flat space-time are evaded by introducing a negative cosmological constant $\Lambda<0$ \cite{Fradkin:1986ka}. In particular, the generators $T_2$ in \eqref{HS:com} form the maximal finite-dimensional subalgebra in the HS algebra, which is the algebra of AdS isometries $o(d-1,2)$; rather than the Poincar\'{e} algebra. Furthermore, a key aspect of this symmetry is that lower-spin fields are sourced by higher-spin fields, which inevitably makes HS interactions with gravity non-minimal and introduces non-locality. Consequently, the HS theory cannot be local in the standard sense; the number of space-time derivatives involved in these interactions increases with the spin, leading to an indefinite growth \cite{Bengtsson:1983pd, Berends:1984rq, Fradkin:1987ks, Fradkin:1991iy}.

The challenge of defining acceptable or minimal non-locality remains unresolved and is currently a topic of intense research. For a list of references on this subject, see e.g., \cite{Kirakosiants:2025gpd}. In \cite{Gelfond:2018vmi}, it was conjectured that HS interactions in four dimensions are all-order spin-local, meaning that for a given set of spins, the HS vertex contains only a finite number of derivatives. This conjecture has been positively tested in a few orders using the Vasiliev equations \cite{Vasiliev:1992av} in \cite{Didenko:2018fgx, Didenko:2019xzz, Didenko:2020bxd, Gelfond:2021two}, where the respective vertices were identified.  In \cite{Didenko:2022qga} (see also \cite{Korybut:2025vdn}), an all-order proof was provided for the (anti)holomorphic sector of the theory, using the so-called irregular form of the Vasiliev equations \cite{Didenko:2026nag}. This gives us access to the complete (anti)holomorphic HS interactions in a local form.  

In this paper, we examine the holomorphic sector of HS theory by analyzing its vertices, as specified in \cite{Didenko:2024zpd}. Our main observation is straightforward: the self-dual HS vertices terminate when HS potentials are set to zero for all spins $s>s_0$. From the discussion above, it is easy to suggest that such truncation is not consistent. In other words, the resulting truncated equations might either lack solutions or only hold true for specific field configurations \footnote{In broader terms, the issue revolves around the field spectrum that allows for nonlinear interactions. If we remove certain fields from the spectrum in the original system, they will reappear in the consistency analysis.}. Surprisingly, when we set  $s_0=2$, we find that the truncated self-dual equations remain consistent. They describe interacting fields of spin zero, as well as fields with negative helicities  $\gl = -1$  and  $\gl = -2$. In other words, a scalar interacting with self-dual electrodynamics and self-dual gravity forms a closed dynamical subsystem in HS theory. Furthermore, focusing on the helicity $\gl=-2$, the corresponding equations describe self-dual gravity. 
As a result, we show that self-dual gravity remains rigid in HS theory, with the corresponding HS vertex offering an unfolded representation of this relationship. Remarkably, its expression involves the Moyal star product, which encodes the HS algebra. Thus, it is particularly interesting to observe that self-dual gravity, when formulated in this manner, clearly exhibits structures that arise from HS symmetry. 

Additionally, we have investigated a closed sub-sector of fields with helicities $-2\leq\gl\leq 0$. The corresponding equations of motions are identified. In particular, we find a contribution from self-dual gravity and electrodynamics to a scalar dynamics that has the following schematic form (see Eq. \eqref{eq:0}):
\begin{equation}
    \Box\phi-2\Lambda\phi=(\text{Weyl})^2+(\text{Maxwell})^2\,.
\end{equation}

Let us emphasize that our result does not mean that any self-dual gravity background with a spin one field and a scalar field qualifies as a solution in HS theory. In fact, self-dual gravity appears to be embedded in a triangular way: it interacts with itself while also serving as a source for all helicities $\gl>-2$. At the same time, no fields with $\gl>-2$ act as sources for gravity.  

\section{Self-dual gravity}
Four-dimensional field theories, when expressed using spinors, can be naturally divided into three parts: the holomorphic part, the antiholomorphic part, and a mixed part that arises from interactions of these two. It is important to note that neither the holomorphic nor the antiholomorphic part is real in Minkowski spacetime. When we turn off the holomorphic part, we are left with a theory of complex fields, which we will refer to as self-dual. 

In the context of gravity, the Weyl tensor is a field that allows for this type of splitting. By introducing two-component $sl(2,C)$ spinor indices $\al, \gb=1,2$ and $\dal, \dgb=1,2$, we can describe the gravitational Weyl tensor using its totally symmetric in spinor indices (anti)self-dual components

\begin{equation}
    {\text{Weyl}}=C_{\al\gb\gga\gd}\oplus\bar{C}_{\dal\dgb\dgga\dot\gd}\,.
\end{equation}
Our convention is to set 
\begin{equation}\label{SD:def}
    C_{\al\gb\gga\gd}=0
\end{equation}
in the self-dual case. Gravitational theory subject to \eqref{SD:def} is called self-dual. It reveals a number of appealing features, such as quantum consistency (one-loop exact). It also describes gravitational instantons, and it is integrable \cite{Plebanski:1975wn}; see also \cite{Chacon:2020fmr} for a recent account. For a comprehensive review and references, we refer to \cite{Krasnov:2016emc}. 

The Cartan formulation of self-dual gravity amounts to the following structure equations
\begin{subequations}\label{GR:Cartan}
    \begin{align}
        &\dr_x\go_{\al\gb}+\go_{\al\gga}\go^{\gga}{}_{\gb}{+}\Lambda\mathbf{e}_{\al}{}^{\dgb}\mathbf{e}_{\gb\dgb}=0\,,\label{GR:SD}\\
        &\dr_x\bar\go_{\dal\dgb}+\bar\go_{\dal\dgga}\bar\go^{\dgga}{}_{\dgb}{+}\Lambda\mathbf{e}^{\gb}{}_{\dal}\mathbf{e}_{\gb\dgb}=\mathbf{e}_{\rho}{}^{\dot\gga}\mathbf{e}^{\rho\dot\gd}\bar{C}_{\dal\dgb\dgga\dot\gd}\,,\label{GR:dWeyl}\\
        &\dr_x\mathbf{e}_{\al\dal}+\go_{\al\gb}\mathbf{e}^{\gb}{}_{\dal}+\bar\go_{\dal\dgb}\mathbf{e}_{\al}{}^{\dgb}=0\,,\label{GR:no torsion}
    \end{align}
\end{subequations}
where 1-forms $\go_{\al\gb}=\go_{\gb\al}$ and $\bar\go_{\dal\dgb}=\bar\go_{\dgb\dal}$ are (anti)selfdual components of the Lorentz connection, $\mathbf{e}_{\al\dgb}=\mathbf{e}_{\al\dgb}^a\dr x_a$ is the vierbein, while $\Lambda$ is the cosmological constant that we assume to be negative $\Lambda<0$.

Equations \eqref{GR:Cartan} express the zero torsion condition via \eqref{GR:no torsion} and self-duality via \eqref{GR:SD} and \eqref{GR:dWeyl}. These are first order differential conditions, which can be completed into the {\it unfolded} form (see \cite{Vasiliev:1988sa}; for the list of recent applications, refer to \cite{Misuna:2022cma, Misuna:2024ccj, Misuna:2024dlx, Misuna:2026bhy, Ammon:2022vjr, Tarusov:2023rad, Boulanger:2024lwk, Tatarenko:2025krq, Iazeolla:2025btr}), provided one resolves the integrability condition $\dr_x^2=0$ through a set of first order equations on the Weyl tensor $\bar{C}_{\dal\dgb\dgga\dot\gd}$ and its on-shell descendants. 

The equations that describe the self-dual gravity Weyl module with a nonzero cosmological constant, along with Eqs. \eqref{GR:Cartan}, as we will show, appear to follow directly from HS theory. They take the following form:
\begin{align}\label{GR:Vas}
    DC_{\gb(m), \dal(m+4)}=&-i\mathbf{e}^{\gga\dgga}C_{\gga\gb(m), \dgga\dal(m+4)}+i m(m+4)\sqrt{-\Lambda}\,\e_{\gb\dal}C_{\gb(m-1), \dal(m+3)}\\
    &+2i\sum_{k=0}^{m-1}\frac{(m+4)!(m-k)}{(m-k+2)!(k+2)!(m+1)}\mathbf{e}_{\gb}{}^{\dgga}C_{\gb(k), \dal(k+2)\dgga(2)}C_{\gb(m-k-1), \dal(m+2-k)}{}^{\dgga}\,,\nonumber
\end{align}
where $D$ is the Lorentz covariant differential, the action of which is specified in the vierbein postulate \eqref{GR:no torsion}, e.g., 
\begin{equation}
    DA_{\al\dal}:=\dr A_{\al\dal}+\go_{\al\gb}A^{\gb}{}_{\dal}+\bar\go_{\dal\dgb}A_{\al}{}^{\dgb}\,.
\end{equation}
We also adopt the standard HS convention, where repeated indices imply symmetrization; for example, $\bar C_{\dal(4)}=\bar C_{\dal_1\dal_2\dal_3\dal_4}$. In the case of pure gravity with $\Lambda=0$ and no higher spins involved, the equations \eqref{GR:Vas} were derived quite some time ago in \cite{Vasiliev:1989xz}.
The infinite set of fields $C_{\al(m), \dal(m+4)}$, where $m>0$ is defined in terms of on-shell derivatives of the self-dual Weyl tensor  $\bar C_{\dal(4)}$ $(m=0)$ by means of Eqs. \eqref{GR:Vas}. Consequently, these fields are auxiliary. 

Equations in \eqref{GR:Vas} are consistent in the sense of $\dr_x^2=0$. A notable feature  is that the expansion in terms of $C$ terminates at the quadratic order. While one might not expect the same behavior from full Einstein gravity, it was emphasized in \cite{Vasiliev:1989xz} that there is a simple kinematic argument why the quadratic approximation is exact in the self-dual case\footnote{The lack of higher-order corrections has to do with a particular index contraction of the vierbein in equation \eqref{GR:Vas}. These higher-order terms could arise if the system is expressed in an improper field frame. We would like to thank M.A. Vasiliev for discussing this point with us.}.

\section{Higher spins in four dimensions}
Higher-spin theory is most developed in lower dimensions, particularly in three\footnote{See also \cite{Blencowe:1988gj, Campoleoni:2010zq, Henneaux:2010xg} for various approaches in three dimensions and \cite{Fredenhagen:2024lps} for recent progress.} dimensions \cite{Prokushkin:1998bq} and four \cite{Vasiliev:1992av}, due to the availability of spinorial isomorphisms in these dimensions. In four dimensions, the unfolded form of the classical HS equations of motion extends the pure gravity case discussed earlier. This is achieved by introducing a set of fields organized into generating functions, utilizing auxiliary variables $y_{\al}$ and $\bar y_{\dal}$:
\begin{equation}\label{omega}
\go(y,\bar{y}|x)=\sum_{m,n=0}^{\infty}\frac{1}{m!n!}\omega_{\alpha_1 \ldots \alpha_m,\dot{\alpha}_1 \ldots \dot{\alpha}_n} y^{\alpha_1}\ldots y^{\alpha_m}\bar{y}^{\dot{\alpha}_1}\ldots \bar{y}^{\dot{\alpha}_n}\,,
\end{equation}
where $\go_{\al(m), \dal(n)}$ is a set of space-time 1-forms. For the case where $m+n=2$, the fields $\go_{\al\gb}$, ${\go}_{\dal\dgb}$, and $\go_{\al, \dgb}$ represent the frame fields of Cartan gravity, as described in Eq. \eqref{GR:Cartan} with $\go_{\al, \dgb}:=\e_{\al\dgb}$.

More generally, the condition 
\begin{equation}
    m+n=2(s-1)
\end{equation}
identifies a set of 1-forms that describe a single spin $s$  gauge field. In contrast, Eq. \eqref{omega} represents an infinite sum of all gauge fields with $s\geq 1$. 

The HS counterparts of the gravitational Weyl tensors  $C_{\al(4)}$ and $\bar C_{\dal(4)}$, along with their descendants specified in the self-dual case in Eq.\eqref{GR:Vas}, are packed in the generating function:
\begin{equation}\label{def: C}
C(y,\bar{y}|x)=\sum_{m,n=0}^{\infty}\frac{1}{m! n!}C_{\alpha_1 \ldots \alpha_m,\dot{\alpha}_1\ldots \dot{\alpha}_n} y^{\alpha_1}\ldots y^{\alpha_m}\bar{y}^{\dot{\alpha}_1}\ldots \bar{y}^{\dot{\alpha}_n}\,.
\end{equation}
the components that satisfy the condition
\begin{equation}\label{s:Weyl}
    m-n=2\lambda
\end{equation}
correspond to a primary helicity $\lambda$ Weyl tensor, which emerges when either  $m=0$ or $n=0$. For example, the component $\bar C_{\dal(4)}$ describes the helicity $\gl=-2$ self-dual Weyl tensor. Additionally, its descendants emerge when  $m\cdot n\neq 0$. As a result, the field $C$ in \eqref{def: C} parametrizes matter fields of $s=0$ and $s=1/2$ and a set of HS Weyl tensors when $s\geq 1$. 

HS equations of motion have the following schematic form \cite{Vasiliev:1999ba}:
\begin{subequations}\label{unfld}
     \begin{align}
        &\dr_x\go=\sum_{n=0}^{\infty}\sum_{i+j=n}\eta^i\bar\eta^j\mathcal{V}_{i,j}(\go, \go, C^n)\,,\label{unfld:w}\\
        &\dr_x C=\sum_{n=0}^{\infty}\sum_{i+j=n}\eta^i\bar\eta^j\Upsilon_{i,j}(\go, C^{n+1})\,,\label{unfld:C}
    \end{align}
    \end{subequations}
where $\mathcal{V}$ and $\Upsilon$ are the so-called interaction vertices, while the only free parameter in the theory is the complex phase $\eta$ and its conjugate $\bar\eta$. The lowest order vertices are defined by the HS algebra represented through the Moyal star product  
\begin{equation}\label{Moyal}
f(y,\bar{y})\ast g(y,\bar{y})=f(y,\bar{y})\exp\left\{i\epsilon^{\alpha \beta} \frac{\overleftarrow{\partial}}{\partial y^\alpha}\frac{\overrightarrow{\partial}}{\partial y^\beta}+i\bar{\epsilon}^{\dot{\alpha}\dot{\beta}}\frac{\overleftarrow{\partial}}{\partial \bar{y}^{\dot{\alpha}}}\frac{\overrightarrow{\partial}}{\partial \bar{y}^{\dot{\beta}}}\right\}g(y,\bar{y})\,.
\end{equation}
Specifically, 
\begin{subequations}\label{initio}
    \begin{align}
        &\mathcal{V}_{0,0}=-\go*\go\,,\label{V0}\\
        &\Upsilon_{0,0}=-\go*C+C*\pi[\go]\,,\label{ups0}
    \end{align}
\end{subequations}
where $\pi$ is the following automorphism of the HS algebra:
\begin{equation}\label{pi}
    \pi[f(y, \bar y)]=f(-y, \bar y)\,.
\end{equation}
The automorphism kinematically constrains Eqs. \eqref{unfld}, leaving them consistent, provided the system is bosonic\footnote{The supersymmetric version of \eqref{unfld} includes fermions; see \cite{Vasiliev:1992av}.}, which implies 
\begin{equation}
    \go(y, -\bar y)=\go(-y, \bar y)\,,\qquad C(y, -\bar y)=C(-y, \bar y)\,.
\end{equation}
The explicit form of the vertices in \eqref{unfld} that satisfy the locality criterion is still unknown. However, the reconstruction of these vertices can be systematically approached using Vasiliev's equations \cite{Vasiliev:1992av}. Recently, the vertices $\mathcal{V}_{j,0}$ and $\Upsilon_{j,0}$ have been identified for all $j\geq 0$ in \cite{Didenko:2024zpd} (see also \cite{Korybut:2025vdn}, \cite{Sharapov:2022nps}) through the so-called irregular formulation of Vasiliev theory \cite{Didenko:2022qga, Didenko:2026nag}. These vertices correspond to the holomorphic or self-dual sector\footnote{Vasiliev's theory is  (HS) background independent and shares a similar spirit with Einstein's gravity in this regard. There is also a perturbative light-cone approach pioneered in \cite{Bengtsson:1983pg} and further developed by Metsaev \cite{Metsaev:1991mt, Metsaev:1993ap}. In the self-dual case, this approach works in flat spacetime. In recent years, it has seen further advancements, such as integrability, color-kinematic duality, and double copy \cite{Ponomarev:2017nrr}. Additionally, a classification program for flat self-dual higher-spin theories was initiated in \cite{Serrani:2025owx}.} of the theory where, for instance, $\eta=0$. In this case, we can rescale $\bar\eta=1$ and express the respective self-dual equations of motion as follows:
\begin{subequations}\label{HS:SD}
\begin{align}
&\dr_x \omega +\omega\ast \omega = \mathcal{V}(\go,\go,C)+...,\\
&\dr_x C+\go\ast C-C\ast \pi[\go]=\Upsilon(\go,C,C)+...\,.
\end{align}
\end{subequations} 
In these equations, the vertices $\mathcal{V}$ and $\Upsilon$ are explicitly defined. While their all-order form is not crucial for our subsequent discussion, we do require that only a limited number of the infinitely many vertices remain nonzero when $\go$ is a polynomial. That this is indeed the case was noted in \cite{Didenko:2019xzz} for lower orders and holds true for all orders \cite{Didenko:2022qga, Didenko:2024zpd}. Specifically, according to the explicit expressions from \cite{Didenko:2024zpd}, if we assume that $\go$ is a polynomial given by
\begin{equation}\label{om:pol}
    \go(y, \bar y)=\sum_{m+n\leq 2(s_0-1)}\frac{1}{m!n!}\go_{\al(m), \dal(n)}(y^{\al})^m(\bar y^{\dal})^n\,,
\end{equation}    
where $s_0$ is a fixed value of the highest  spin in $\go$, we then observe using the analysis from \cite{Didenko:2025xca} that   
\begin{equation}\label{trunc:s}    
    \mathcal{V}(\go, \go, \underbrace{C\dots C}_{k\geq 2(s_0-1)})=\Upsilon(\go, \underbrace{C\dots C}_{k\geq 2s_0-1})=0\,. 
\end{equation}
This indicates that for a polynomial $\go$, an infinite series of HS vertices effectively terminates. As previously discussed, the HS algebra does not support a finite spin truncation, so we do not expect such a truncation to be generally consistent. Nevertheless, we argue that for $s_0=2$, the self-dual truncation turns out to be consistent.

\section{Truncation}

Consider the self-dual system \eqref{HS:SD} and let us assume the 1-form $\go$ is restricted only to spins one and two, i.e., we set $s_0=2$ in \eqref{om:pol}. This implies 
\begin{equation}\label{trunk}
\omega(y,\bar{y}|x)\Big|_{s\leq 2}:=\Omega(y,\bar{y}|x):= i\left(A-\frac{1}{4}\left(\omega^{\alpha \beta} y_\alpha y_\beta+\bar{\omega}^{\dot{\alpha}\dot{\beta}}\bar{y}_{\dot{\alpha}}\bar{y}_{\dot{\beta}}+2\mathbf{e}^{\alpha \dot{\beta}} y_\alpha \bar{y}_{\dot{\beta}}\right)\right)\,,
\end{equation}
where the $y$ independent part $A$ is the spin $s=1$ Maxwell potential, while $\go_{\al\gb}$, $\bar\go_{\dal\dgb}$, and $\e_{\al\dgb}$ are the standard Cartan gravity frame fields. 

Now, the self-dual HS equations \eqref{HS:SD} supplemented  by the condition \eqref{trunk} take the following truncated form:
\begin{subequations}\label{HS:trunc}
\begin{align}
&\dr_x \Omega +\Omega\ast \Omega = \mathcal{V}(\Omega,\Omega,C)\,,\label{eq:w}\\
&\dr_x C+\Omega\ast C-C\ast \pi[\Omega]=\Upsilon(\Omega,C,C)\,.\label{eq:C}
\end{align}
\end{subequations} 
By definition, Eqs. \eqref{HS:trunc} follow from the complete consistent system \eqref{HS:SD}. This means that Eq.  \eqref{eq:w} is consistent up to order $O(C)$, while Eq. \eqref{eq:C} is consistent up to order $O(C^2)$. Inspecting integrability $\dr_x^2=0$, it is easy to derive the consistency condition of the full system \eqref{HS:trunc}, which has the following form:
\begin{subequations}\label{eqs:cons}
    \begin{align}
    &\mathcal{V}(\mathcal{V}(\Omega, \Omega, C), \Omega, C)-\mathcal{V}(\Omega, \mathcal{V}(\Omega, \Omega, C), C)+\mathcal{V}(\Omega, \Omega, \Upsilon(\Omega, C, C))=0\,,\\
    &\Upsilon(\mathcal{V}(\Omega, \Omega, C), C, C)-\Upsilon(\Omega, \Upsilon(\Omega, C, C), C)-\Upsilon(\Omega, C, \Upsilon(\Omega, C, C))=0\,. \label{appendix}
\end{align}
\end{subequations}
It is important to note that Eqs. \eqref{eqs:cons} are not part of the complete consistency requirements for HS interaction. As a result, they may not necessarily be fulfilled. To verify whether the equations in \eqref{eqs:cons} hold, we need the explicit form of the vertices. The vertex from \eqref{eq:w} corresponds to the well-known self-dual part of the central on-mass-shell theorem \cite{Vasiliev:1988sa}. For the vertex in \eqref{eq:C}, which first appeared in \cite{Vasiliev:2016xui}, we will use the presentation provided in \cite{Didenko:2025xca}\footnote{Unlike \cite{Didenko:2025xca}, we adopt the reality condition $C^\dagger=\pi (C)$. This choice accounts for the difference in normalization in \eqref{SG0}.}. Explicitly,
\begin{subequations}\label{eqs:SD}
\begin{align}\label{SG1}
&\dr_x \Omega+\Omega\ast \Omega=-\frac{i}{2}\mathbf{e}_{\alpha} {}^{\dot{\alpha}} \mathbf{e}^{\alpha \dot{\beta}}\frac{\partial^2}{\partial \bar{y}^{\dot{\alpha}}\, \partial \bar{y}^{\dot{\beta}}}C(0,\bar{y})\,,\\
\label{SG0}
&\dr_x C+\Omega\ast C-C\ast \pi[\Omega]= -\mathbf{e}^{\beta \dot{\beta}} y_\beta \int_0^1 d\tau\, (1-\tau)\left[\frac{\partial}{\partial \bar{y}^{\dot{\beta}}} C\big(\tau y,\bar{y}\big),C((1-\tau)y,\bar{y})\right]_{\bar{\ast}}\,,
\end{align}
\end{subequations}
where $\bar *$ is a part of the Moyal product \eqref{Moyal} that acts only on the variables $\bar y$. The main result of this letter is the statement that this system is consistent. We will prove it in the Appendix. An interesting feature of this system is that fields with lower helicities, specifically those in the range $-2\leq\gl\leq 0$, source each other as well as fields with positive helicities. In contrast, fields with positive helicities $\gl_+>0$ do not source the lower helicity fields. This means that the interactions among the fields in the range $-2\leq\gl\leq 0$ form a triangular structure within. The interaction can be summarized in the table below.
\begin{table}[h]
\centering
\begin{tabular}{|c|c|c|c|c|}
\hline 
 & $0$ & $1_-$  & $2_-$ & $\gl_+$\\ 
\hline 
$0$  & $\gl_+$ & $\gl_+$  & $0$+ $\gl_+$ & $\gl_+$  \\ 
\hline 
$1_-$ & $\gl_+$  & $0$+$\gl_+$  & $1_-$+$\gl_+$ & $\gl_+$   \\ 
\hline 
$2_-$ & $0$+$\gl_+$ &  $1_-$+$\gl_+$ & $0$+$2_-$+$\gl_+$ & $\gl_+$ \\ 
\hline 
$\gl_+$ & $\gl_+$ & $\gl_+$ & $\gl_+$ & $\gl_+$\\
\hline
\end{tabular} 
\caption{Columns and rows indicate the helicities $\gl=-2, -1, 0, \gl_+$ entering field  $C$ on the right of \eqref{SG0}, while the intersection provides their contribution in interaction.}
\end{table}
Assuming consistency, here we provide a brief summary of its content:
\begin{itemize}
    \item {\bf Mostly positive helicities $\gl\geq -2$.} From \eqref{SG1} it follows that the negative helicity $\gl<-2$ HS Weyl tensors should be set to zero
    \begin{equation}\label{no hs}
        \bar C_{\dal_1\dots\dal_{2s}}=0\,,\qquad s>2\,.
    \end{equation}
    This follows from the fact that (i) only negative helicities contribute to the right-hand side of \eqref{SG1}, and (ii) the left-hand side being at most quadratic in $\bar y$ implies \eqref{no hs}. Additionally, it is straightforward to observe, using \eqref{Moyal}, that the right-hand side of \eqref{SG0} does not contribute to helicities $\gl < -2$ when both $C$' have helicities $\gl_{1,2}\geq-2$.
    
    \item {\bf Helicity $\gl=-1$.} Using the $Y$-independent component of \eqref{trunk} and \eqref{def: C}, from \eqref{SG1} we have 
    \begin{equation}
         \dr_x A=-\frac{1}{2}\mathbf{e}_{\alpha} {}^{\dot{\alpha}} \mathbf{e}^{\alpha \dot{\beta}}\bar C_{\dal\dgb}\,,
    \end{equation}
    which is simply Maxwell's equation in a curved background, where $\bar{C}_{\dal\dgb}$ is the self-dual Faraday tensor. Its unfolded equations are encoded in \eqref{SG0}, as we can extract them by zooming in the helicity $\gl=-1$ module spanned by the fields $C_{\al(m), \dal(m+2)}$, where $m\geq 0$ in accordance with \eqref{s:Weyl}. Specifically, from \eqref{SG0} we have
    \begin{multline}\label{eq:1}
        DC^{\al(m), \dot{\beta}(m+2)}=i\sqrt{-\Lambda}\, m(m+2) \mathbf{e}^{\alpha \dot{\beta}} C^{\alpha(m-1),\dot{\beta}(m+1)}-i \mathbf{e}_{\sigma \dot{\gamma}} C^{\alpha(m)\sigma,\dot{\beta}(m+2)\dot{\gamma}}+\\
        +2i\mathbf{e}^\alpha {}_{\dot{\gamma}} \sum_{k=0}^{m-1}\Bigg(\frac{(m-k)\, m!\, (m+2)}{k!\, (m-k+2)!} C^{\alpha(k),\dot{\beta}(k)} {}_{\dot{\sigma}} {}^{\dot{\gamma}} C^{\alpha(m-k-1),\dot{\beta}(m-k+2)\dot{\sigma}}+\\
        +\frac{m!\, (m-2)}{(k+2)!\, (m-k-1)!} C^{\alpha(k),\dot{\beta}(k+2)}{}_{\dot{\sigma}} {}^{\dot{\gamma}} C^{\alpha(m-k-1),\dot{\beta}(m-k) \dot{\sigma}}\Bigg)\,,
    \end{multline}
where, for convenience, we have restored the cosmological constant $\Lambda$, which was set to $\Lambda=-1$ in \eqref{SG0}. Notice that the second and third lines in Eq. \eqref{eq:1} include  contributions from interaction with  the helicity $\gl=-2$ field. Equation \eqref{eq:1} encodes the standard self-dual Maxwell equations in presence of gravity
\begin{equation}
    D_{\al}{}^{\dot\gga}\bar C_{\dot\gga\dal}=0\,,
\end{equation}
which can be obtained by inspecting the $m=0$ component of \eqref{eq:1}, and where we also used $D:=\mathbf{e}^{\al\dgb}D_{\al\dgb}$.
 
    \item {\bf Helicity $\gl=-2$.}
    From the $Y$-dependent part of $\Omega$ in \eqref{trunk}, we obtain Eqs. \eqref{GR:Cartan} of self-dual gravity, where the $\gl=-2$ HS Weyl module $C_{\al(m), \dal(m+4)}$ satisfies \eqref{GR:Vas}. These equations can be easily extracted from \eqref{eqs:SD}, allowing us to reproduce the self-dual gravity unfolded equations obtained in \cite{Vasiliev:1989xz} and generalize them to include a nonzero cosmological constant.

 Curiously, the self-dual gravity vertex \eqref{GR:Vas} acquires a form which contains remnants of the star-product algebra represented by the Moyal product $\bar *$
\begin{equation}
    \Upsilon(\e, \mathcal{C},\mathcal{C})={-}\mathbf{e}^{\beta \dot{\beta}} y_\beta \int_0^1 d\tau\, (1-\tau)\left[\frac{\partial}{\partial \bar{y}^{\dot{\beta}}} \mathcal{C}\big(\tau y,\bar{y}\big),\mathcal{C}((1-\tau)y,\bar{y})\right]_{\bar{\ast}}^{\gl=-2}\,,
\end{equation}
where we denote the self-dual gravity Weyl module as
\begin{equation}
    \mathcal{C}:=\sum_{m=0}^{\infty}\frac{1}{m!(m+4)!}C_{\al(m), \dal(m+4)}(y^{\al})^m(\bar y^{\dal})^{m+4}\,,
\end{equation}
and $[]^{\gl=-2}$ means the projection on the helicity $\gl=-2$ module. It is important to emphasize that the vertex \eqref{SG0} does not include contributions from any helicities other than purely gravitational ones to the helicity $\gl=-2$ sector. As a result, self-dual gravity becomes a fundamental rigid component of the HS theory.

\item {\bf Scalar field.} A scalar field Weyl module is represented by the components $C_{\al(m), \dal(m)}$, where the lowest component $m=0$ is associated with the primary scalar field $\phi(x)=C(0,0|x)$. The unfolded equations extracted from \eqref{SG0} read
\begin{multline}
D C^{\alpha(m),\dot{\beta}(m)}-i{\sqrt{-\Lambda}}\, m^2 \mathbf{e}^{\alpha \dot{\beta}}C^{\alpha(m-1),\dot{\beta}(m-1)}+i \mathbf{e}_{\sigma \dot{\gamma}} C^{\alpha(m)\sigma,\dot{\beta}(m)\dot{\gamma}}=\\
=-\frac{i}{3}\sum_{k=0}^{m-1}\frac{m!}{k!(m-k-1)!(m+1)}\, \mathbf{e}^\alpha {}_{\dot{\gamma}}C^{\alpha(k),\dot{\beta}(k)}{}_{\dot{\sigma}(3)} {}^{\dot{\gamma}}\, C^{\alpha(m-k-1),\dot{\beta}(m-k)\dot{\sigma}(3)}+\\
+2i\sum_{k=0}^{m-1}\frac{m!}{k!(m-k-1)!(m+1)}\, \mathbf{e}^{\alpha \dot{\gamma}}C^{\alpha(k),\dot{\beta}(k)}{}_{\dot{\xi}\dot{\gamma}}\, C^{\alpha(m-k-1),\dot{\beta}(m-k)\dot{\xi}}\\
+2i\sum_{k=0}^{m-1}\frac{(m-k)\, m!}{(k-2)! (m-k+2)! (m+1)} \mathbf{e}^\alpha {}_{\dot{\gamma}}C^{\alpha(k),\dot{\beta}(k-2)}{}_{\dot{\sigma}}{}^{\dot{\gamma}} C^{\alpha(m-k-1),\dot{\beta}(m-k+2)\dot{\sigma}}+\\
+2i\sum_{k=0}^{m-1}\frac{(m-k)\, m!}{(k+2)! (m-k-2)! (m+1)} \mathbf{e}^\alpha {}_{\dot{\gamma}} C^{\alpha(k),\dot{\beta}(k+2)} {}_{\dot{\sigma}} {}^{\dot{\gamma}} C^{\alpha(m-k-1),\dot{\beta}(m-k-2)\dot{\sigma}}\,.
\end{multline}

From the components with $m=0$ and $m=1$, we straightforwardly find that
\begin{equation}\label{eq:0}
    \Box \phi{-2\Lambda}\,\phi=\frac{1}{3}\bar C_{\dal\dgb\dgga\dot\delta}\bar C^{\dal\dgb\dgga\dot\delta} -\bar C_{\dot{\alpha}\dot{\beta}} \bar C^{\dot{\alpha}\dot{\beta}}\,,\qquad \Box:=\frac{1}{2}D_{\al\dal}D^{\al\dal}\,,
\end{equation} 
Notice that the scalar field is sourced by the self-dual Weyl tensor along with the Maxwell tensor. 
\end{itemize}

\section{Conclusion}

We have shown that HS theory allows for a consistent decoupling of all gauge fields with spin $s>2$ within its self-dual sector. The remaining spectrum consists of gauge fields (1-forms) associated with self-dual electrodynamics\footnote{For the colored HS theory, electrodynamics is naturally replaced with self-dual YM.} and gravity, as well as field strengths (0-forms) of all helicities $\gl\geq -2$. The interactions are organized through a universal coupling with a gravitational vierbein.

There is also a small sub-sector of helicities $-2\leq\gl\leq 0$ that is not sourced by other fields, and we have identified the corresponding nonlinear dynamics for this sector. Notably, self-dual gravity represents a rigid component of the theory that remains unaffected by higher-spin interactions. This conclusion is supported by the unfolded equations we derived from the HS vertex, which were previously constructed in \cite{Vasiliev:1989xz} for pure self-dual gravity.  The appearance of this closed sector can be thought of as a result of factoring the ideal related to positive helicities in the $L_{\infty}$ algebra represented by the maps in \eqref{HS:SD}. Considering the observation in \eqref{trunc:s}, it would be interesting to investigate whether \eqref{HS:SD} allows for other finite-spin truncations.

A notable aspect of the observed self-dual gravity equations is that they clearly exhibit a remnant of HS symmetry, which is realized through the holomorphic Moyal product. This perspective puts gravity into the framework of HS interactions, potentially aiding in the search for its AdS/CFT dual along the lines of \cite{Vasiliev:2012vf, Diaz:2024kpr, Diaz:2024iuz}. Additionally, self-dual gravity has recently been shown \cite{Lipstein:2023pih} to admit a form of a nonlinear scalar field. It would be interesting to relate the two pictures. It is also of interest to consider our findings in relation to \cite{Neiman:2024vit}, where self-dual gravity is examined within the context of higher-spin theories.

In deriving the truncated HS system \eqref{eqs:SD}, we used an observation from \cite{Didenko:2019xzz}, which demonstrated that self-dual HS vertices vanish when the gauge fields' 1-forms are limited to lower spins. This raises a natural problem of field configurations supporting such a truncation. Essentially, this issue revolves around analyzing specific algebraic constraints that arise from the requirement for consistency. Notably, when the spins of the 1-forms are restricted to $s\leq 2$, the nonlinear algebraic constraints hold true for all field-strength helicities  $\gl\geq -2$. This happens because the vertex \eqref{SG0} is delicately finetuned. Its integration measure and the dependence on the integration variable $\tau$ are structured in such a way that the cubic algebraic constraint $O(C^3)$ reduces to an integral over a two-dimensional simplex, where the integrand appears to be antisymmetric with respect to the vertices of the simplex. This leads to the integration resulting in zero.

\section*{Acknowledgments}
We would like to thank M.A. Vasiliev for the fruitful discussions. We also appreciate the valuable comments on the draft from I. Faliakhov and D. Ponomarev. V.D. is grateful for the financial support from the Foundation for the Advancement of Theoretical Physics and Mathematics “BASIS”.

\section{Appendix. Consistency}

We verify that constraint \eqref{appendix} indeed holds for a particular expression for the vertex \eqref{SG0}. Note that the first term in \eqref{appendix} is trivial (see \eqref{GR:no torsion}, which is part of Eq. \eqref{SG1}). So we are left with
\begin{multline}
\mathbf{e}^{\beta \dot{\beta}} y_\beta \int_0^1 d\mathcal{T}\, (1-\mathcal{T})\left[\frac{\partial}{\partial \bar{y}^{\dot{\beta}}} \Upsilon(\mathbf{e},C,C)(\mathcal{T}y,\bar{y}),C\big((1-\mathcal{T})y,\bar{y}\big)\right]_{\bar{\ast}}+\\
+i\mathbf{e}^{\beta \dot{\beta}} y_\beta \int_0^1 d\mathcal{T}\, (1-\mathcal{T})\left[\frac{\partial}{\partial \bar{y}^{\dot{\beta}}} C\big(\mathcal{T}y,\bar{y}\big),\Upsilon(\mathbf{e},C,C)((1-\mathcal{T})y,\bar{y})\right]_{\bar{\ast}}
\end{multline}
that, if not zero, breaks consistency. 
More explicitly, this expression looks as follows:
\begin{multline}\label{Pre1}
\mathbf{e}^{\beta \dot{\beta}} y_\beta \int_0^1 d\mathcal{T}\, (1-\mathcal{T})\times\\
\times\Bigg\{\Bigg[\frac{\partial}{\partial \bar{y}^{\dot{\beta}}}\Bigg(i\mathbf{e}^{\alpha\dot{\alpha}}{\mathcal{T}}y_\alpha\int_0^1 d\xi \, (1-\xi)\Bigg[\frac{\partial}{\partial \bar{y}^{\dot{\alpha}}} C\big(\xi \mathcal{T} y,\bar{y}\big),C\big((1-\xi)\mathcal{T}y,\bar{y}\big)\Bigg]_{\bar{\ast}}\Bigg), C\big((1-\mathcal{T})y,\bar{y}\big)\Bigg]_{\bar{\ast}}+\\
+\Bigg[\frac{\partial}{\partial \bar{y}^{\dot{\beta}}} C\big(\mathcal{T}y,\bar{y}\big),i\mathbf{e}^{\alpha\dot{\alpha}}{(1-\mathcal{T})} y_\alpha \int_0^1 d\xi \, (1-\xi)\Bigg[\frac{\partial}{\partial \bar{y}^{\dot{\alpha}}}C\big((1-\mathcal{T})\xi y,\bar{y}\big),C\big((1-\xi)(1-\mathcal{T})y,\bar{y}\big)\Bigg]_{\bar{\ast}}\Bigg]_{\bar{\ast}}\Bigg\}\,.
\end{multline}
Using the identity
\begin{equation}
\mathbf{e}^{\alpha\dot{\alpha}} \mathbf{e}^{\beta \dot{\beta}}=\frac{1}{2}\epsilon^{\alpha \beta}\,\mathbf{e}_\xi {}^{\dot{\alpha}} \mathbf{e}^{\xi\dot{\beta}}+\frac{1}{2}\bar{\epsilon}^{\dot{\alpha}\dot{\beta}}\,\mathbf{e}^\alpha {}_{\dot{\xi}} \mathbf{e}^{\beta \dot{\xi}}\,,
\end{equation}
and that
\begin{equation}
\int_0^1 d\xi \int_0^1 d\mathcal{T} \, f(\xi,\mathcal{T})=\int d^3 \tau_+\, \delta\left(1-\sum_{i=1}^3\tau_i\right)\, \frac{1}{\tau_1+\tau_2} f\left(\frac{\tau_1}{\tau_1+\tau_2},{\tau_1+\tau_2}\right)
\end{equation} 
where the following shorthand notation is used
\begin{equation}
\int d^3 \tau_+:=\int d\tau_1 \, d\tau_2\, d\tau_3\, \theta(\tau_1)\theta(\tau_2)\theta(\tau_3)\,,
\end{equation}
expression \eqref{Pre1} casts into
\begin{multline}\label{Pre2}
\frac{1}{2}\,\mathbf{e}^\alpha {}_{\dot{\xi}} \mathbf{e}^{\beta \dot{\xi}}\, y_\alpha y_\beta \, {\epsilon}^{\dot{\alpha}\dot{\beta}}\int d^3 \tau_+\, \delta\left(1-\sum_{i=1}^3\tau_i\right)\times\\
\times \Bigg\{\frac{\tau_3\tau_2}{\tau_1+\tau_2}\Bigg[\Big[\frac{\partial}{\partial \bar{y}^{\dot{\alpha}}}C(\tau_1 y,\bar{y}),\frac{\partial}{\partial \bar{y}^{\dot{\beta}}} C(\tau_2 y,\bar{y})\Big]_{\bar{\ast}}, C(\tau_3 y,\bar{y})\Bigg]_{\bar{\ast}}+\\
+\tau_3\Bigg[\frac{\partial}{\partial \bar{y}^{\dot{\beta}}} C(\tau_2 y,\bar{y}),\Big[\frac{\partial}{\partial \bar{y}^{\dot{\alpha}}}C(\tau_1 y,\bar{y}),C(\tau_3 y,\bar{y})\Big]_{\bar{\ast}}\Bigg]_{\bar{\ast}}\Bigg\}\,.
\end{multline}
The structure of the double commutator is not relevant. Expanding the double commutator, we notice that some of the terms proportional to $\tau_3$ cancel each other, leaving us with
\begin{multline}
\frac{1}{2}\,\mathbf{e}^\alpha {}_{\dot{\xi}} \mathbf{e}^{\beta \dot{\xi}}\, y_\alpha y_\beta \,{\epsilon}^{\dot{\alpha}\dot{\beta}}\int d^3 \tau_+\, \delta\left(1-\sum_{i=1}^3\tau_i\right)\, \tau_3 \times\\
\times\Bigg\{\frac{\partial}{\partial \bar{y}^{\dot{\alpha}}}C(\tau_1 y,\bar{y})\bar{\ast} C(\tau_3 y,\bar{y})\bar{\ast}\frac{\partial}{\partial \bar{y}^{\dot{\beta}}} C(\tau_2 y,\bar{y})+\frac{\partial}{\partial \bar{y}^{\dot{\beta}}} C(\tau_2 y,\bar{y})\bar{\ast}C(\tau_3 y,\bar{y})\bar{\ast}\frac{\partial}{\partial \bar{y}^{\dot{\alpha}}}C(\tau_1 y,\bar{y})  \Bigg\}\,.
\end{multline}
The expression in brackets $\{\}$ being symmetric with respect to $\dot{\alpha}$ and $\dot{\beta}$ is contracted with ${\epsilon}^{\dot{\alpha}\dot{\beta}}$ and therefore vanishes. Thus, considering the remaining part of \eqref{Pre2}, we are left with two types of terms:
\begin{itemize}
\item Terms $\bullet \bar{\ast} C(\tau_3 y,\bar{y})$
\begin{multline}
\frac{1}{2}\,\mathbf{e}^\alpha {}_{\dot{\xi}} \mathbf{e}^{\beta \dot{\xi}}\, y_\alpha y_\beta \, {\epsilon}^{\dot{\alpha}\dot{\beta}}\int d^3 \tau_+\, \delta\left(1-\sum_{i=1}^3\tau_i\right)\times\\
\times\left(-\tau_3+\frac{\tau_3 \tau_2}{\tau_1+\tau_2}+\frac{\tau_3\tau_1}{\tau_1+\tau_2}\right)\frac{\partial}{\partial \bar{y}^{\dot{\alpha}}}C(\tau_1 y,\bar{y})\bar{\ast}\frac{\partial}{\partial \bar{y}^{\dot{\beta}}} C(\tau_2 y,\bar{y})\bar{\ast} C(\tau_3 y,\bar{y})\equiv 0\,.
\end{multline}
\item Terms $C(\tau_3 y,\bar{y})\bar{\ast} \bullet$
\begin{multline}
\frac{1}{2}\,\mathbf{e}^\alpha {}_{\dot{\xi}} \mathbf{e}^{\beta \dot{\xi}}\, y_\alpha y_\beta \, {\epsilon}^{\dot{\alpha}\dot{\beta}}\int d^3 \tau_+\, \delta\left(1-\sum_{i=1}^3\tau_i\right)\times\\
\times\left(\tau_3-\frac{\tau_3 \tau_2}{\tau_1+\tau_2}-\frac{\tau_3\tau_1}{\tau_1+\tau_2}\right)C(\tau_3 y,\bar{y})\bar{\ast}\frac{\partial}{\partial \bar{y}^{\dot{\alpha}}}C(\tau_1 y,\bar{y})\bar{\ast}\frac{\partial}{\partial \bar{y}^{\dot{\beta}}} C(\tau_2 y,\bar{y})\equiv 0\,.
\end{multline}
\end{itemize}
They both vanish identically on their own. Consequently, \eqref{appendix} holds.

\end{document}